\documentclass{elsart1p}
\pdfoutput=1
\usepackage{graphics}
\usepackage{graphicx}
\usepackage{epsfig}
\usepackage{amssymb}

\begin{document}

\begin{frontmatter}
\title{Chiral SU(3) Dynamics and \\Antikaon-Nuclear Quasibound States}
\author{W. Weise  and R. H\"artle}
\address{Physik-Department, Technische Universit\"at M\"unchen, D-85747 Garching, Germany}
\begin{abstract}
Recent developments are summarised concerning low-energy $\bar{K}N$ interactions as they relate to the possible existence of antikaon-nuclear quasibound states. An exploratory study of antikaons bound to finite nuclei is performed, with emphasis on the evolution of such states from light to heavy nuclei (A = 16 - 208). The energy dependent, driving attractive $\bar{K}N$ interactions are constructed using the $s$-wave coupled-channel amplitudes involving the $\Lambda(1405)$ and resulting from chiral $SU(3)$ dynamics, plus  $p$-wave amplitudes dominated by the $\Sigma(1385)$. Effects of Pauli and short-range correlations are discussed. The decay width induced by $K^-NN$ two-body absorption is estimated and found to be substantial. It is concluded that $\bar{K}$-nuclear quasibound states can possibly exist with binding energies ranging from 60 to 100 MeV, but with short life times corresponding to decay widths of the similar magnitudes. 
\end{abstract}
\end{frontmatter}
\section{Introduction and outlook}
\label{intro}
The low-energy interactions of kaons with nuclear systems are governed by the spontaneous and explicit breaking of chiral $SU(3)_L\times SU(3)_R$ symmetry in QCD. Spontaneous chiral symmetry breaking assigns the role of Goldstone bosons to the octet of light pseudoscalar mesons. Explicit symmetry breaking by the small but non-vanishing masses of the light quarks shifts the masses of these mesons to their observed positions. The strange quark mass, $m_s\sim 0.1$ GeV, can still (with caution) be considered small compared to the characteristic scale of spontaneous chiral symmetry breaking, $\Lambda_\chi = 4\pi f \sim 1$ GeV, expressed in terms of the pseudoscalar decay constant, $f\simeq 0.09$ GeV. 

Given this symmetry breaking pattern of low-energy QCD, the leading (Tomozawa-Weinberg) interactions of kaons and antikaons with nucleons are determined unambiguously. In particular, the driving $\bar{K}N$ interaction in the isospin $I=0$ channel is strongly attractive around threshold, $\omega \simeq m_K$. Early discussions of kaon condensation in dense matter \cite{KapNel} were based on this observation which, ever since, has given rise to speculations about  the possible existence of antikaon-nuclear bound states \cite{Wy86}.

Subsequent studies of antikaons in neutron star matter \cite{BLRT94} and dense nuclear matter \cite{WKW96,WRW97} came to the qualitative conclusion that in-medium $K^-$ modes are expected to experience strong attraction, corresponding to binding energies of the order of 100 MeV at normal nuclear densities. Under such conditions the primary decay channel of the in-medium antikaonic mode, $\bar{K}N \rightarrow \pi\Sigma$, is expected to be suppressed \cite{WKW96}, and consequently its width should be reduced. However, these early considerations did not yet take into account the two-body absorption channel $\bar{K}NN\rightarrow YN$ with $Y = \Lambda, \Sigma$. This process contributes a term to the total width that grows proportional to the square of the baryon density.  

The recent revival of this theme, now focused on deeply bound states of kaonic nuclei, 
was prompted by Ref.\cite{AY02,YA02}. A simple phenomenological $\bar{K}N$ potential model was used there to compute bound states of few-body systems such as $K^-pp$, $K^-ppn$ and $K^-pnn$. It was again argued that if the binding is sufficiently strong to fall below the $\bar{K}N\rightarrow \pi\Sigma$ threshold, such states could be narrow and long-lived. An experiment performed at KEK with stopped $K^-$ on $^4$He \cite{Su04} seemed indeed to indicate deeply bound narrow structures with widths 
$\Gamma < 20$ MeV.  However,  the subsequent repetition of this experiment with better statistics \cite{Iwa06} did not confirm the previously published results. The FINUDA measurements with stopped $K^-$ on $^{6,7}$Li and $^{12}$C targets \cite{Ag05} suggested an interpretation in terms of quasibound $K^-pp$ clusters with binding energy $B(K^-pp) = (115 \pm 9)$ MeV and width $\Gamma = (67\pm 16)$ MeV. However, this interpretation was criticized in Ref.\cite{OT06} with the argument that the observed spectrum may be explained by final state interactions of the produced $\Lambda p$
pairs. FINUDA \cite{Ag07} has also reported a bump in the $\Lambda d$ invariant mass spectrum following stopped $K^-$ in $^6 Li$, which is being discussed as a signature of a $K^-ppn$ cluster with $B(K^-ppn) = (58\pm 6)$ MeV and $\Gamma = (37\pm 14)$ MeV. Another line of experimental studies focuses on the invariant mass spectroscopy of $\Lambda p$
pairs produced in heavy-ion collisions at GSI and analyzed with the FOPI detector \cite{GSI06}. 

Calculations investigating the possibilty of quasibound antikaon-nuclear few-body systems have been performed with improved $NN$ and $\bar{K}N$ interactions, using either Faddeev \cite{SGM06,IS07} or variational approaches \cite{DW07,YA07}. In such calculations, the prototype $K^-pp$ system is found to be quasibound with binding energies in the range $50 - 80$ MeV, above the $\pi\Sigma$ threshold and consequently with a relatively large width. An extremely dense and compact system as suggested in \cite{AY02} is excluded once a nucleon-nucleon interaction with realistic short-range repulsion is introduced. Further developments using $\bar{K}N$ interactions explicitly constructed from chiral $SU(3)$ coupled-channel dynamics \cite{HW07} and implemented in a variational calculation \cite{DHW08} suggest lower  $K^-pp$ binding energies, around 20 MeV, again with large widths. On the other hand, when a separable approximation is applied to a similar chiral $SU(3)$ based interaction in a Faddeev calculation \cite{IS07}, stronger binding is found ($B\sim 80$ MeV with a width of comparable magnitude). 

The broad range of computed binding energies for the $K^-pp$ three-body system indicates the limited predictive power of all such investigations, given that the energy region of the $\bar{K}N$ interaction relevant for deeply bound states is quite far below the $\bar{K}N$ threshold. Ambiguities in performing such extrapolations require a careful assessment \cite{HW07}. While these issues remain yet unsettled in the absence of conclusive empirical evidence, the experimental searches for antikaonic nuclei continue vigorously. 

Studies of antikaonic nuclei with larger mass numbers have been performed in Refs.\cite{MFG06,GF07}, based on either a $K^-$-nuclear interaction constrained empirically by kaonic atoms \cite{FG07}, or a relativistic nuclear (sigma-omega) mean field model using a schematic $\bar{K}N$ ineraction, in which the nuclear density distribution dynamically adjusts to the attractive interaction provided by the kaon field. 

Our present exploratory study is complementary to the one in \cite{MFG06} in several respects. The strong energy dependence of the s-wave $\bar{K}N$ amplitudes is taken into account guided by recent results from chiral $SU(3)$ dynamics with coupled channels \cite{BNW05}. Effects of p-wave interactions involving the $\Sigma(1385)$ are incorporated. Pauli corrections, charge exchange effects and short-range $NN$ correlations are implemented as in \cite{WRW97}. We do not explicitly calculate, however, the dynamical response of the nuclear core distribution to the presence of the interacting kaon field, as it has been done in \cite{MFG06}. Instead we investigate the variation of results when increasing the nuclear central density, $\rho_0$, beyond its standard value $\rho_0 \simeq 0.17$ fm$^{-3}$. We also estimate two-body absorptive contributions to the widths of antikaonic quasibound states.

\section{Chiral SU(3) dynamics and low-energy $\bar{K}N$ interactions}
\label{chiral}
Chiral perturbation theory (ChPT) as a systematic expansion in small momenta and quark masses
is limited to low-energy processes with light quarks. It is a valid question to what extent 
the generalisation of ChPT including strangeness can be made to work. The $\bar{K} N$ channel is of particular interest in this context, as a testing ground for chiral $SU(3)$ symmetry in QCD and for the role of explicit chiral symmetry breaking by the strange quark mass. However, any perturbative approach breaks down in the vicinity of resonances. 
In the $K^- p$ channel, for example, the existence of the $\Lambda(1405)$ resonance 
just below the $K^- p$ threshold renders $SU(3)$ ChPT
inapplicable. At this point the combination with non-perturbative coupled-channels
techniques has proven useful, by generating the $\Lambda(1405)$ dynamically
as an $I=0$ $\bar{K} N$ quasibound state embedded in the resonant $\pi \Sigma$
continuum \cite{KSW95}. Coupled-channels methods combined with chiral $SU(3)$ dynamics have subsequently been applied to a variety of meson-baryon scattering processes with quite some success \cite{KWW97}. Recent updates can be found in Refs.\cite{BNW05,BMN06}. 

The starting point is the chiral $SU(3)_L \times SU(3)_R$ meson-baryon effective Lagrangian. Its leading order terms include the octet of pseudoscalar Goldstone bosons ($\pi, K, \bar{K}, \eta$) and their interactions. Symmetry breaking mass terms introduce the light quark masses $m_u, m_d$ and the mass of the strange quark, $m_s$. The pseudoscalar mesons interact with the baryon octet ($p, n, \Lambda, \Sigma, \Xi$) through vector and axial vector combinations of their
fields. At this stage the parameters of the theory, apart from the pseudoscalar meson decay constant $f \simeq 0.1$ GeV, are the $SU(3)$ baryon axial vector coupling constants
$D \simeq 0.80$ and $F \simeq 0.47$ which add up to $D + F = g_A = 1.27$. At next-to-leading order, seven additional constants enter in $s$-wave channels, three of which are constrained by mass splittings in the baryon octet and the remaining four need to be fixed by comparison with low-energy scattering data.

\subsection{Coupled channels} 
Meson-baryon scattering amplitudes based on the $SU(3)$ effective Lagrangian involve coupled channels for each set of quantum numbers. For example, The $K^- p$ system in the isospin $I=0$ sector couples strongly to the $\pi\Sigma$ channel. 
Consider the $T$ matrix ${\bf T}_{ij}(p, p')$ connecting meson-baryon channels $i$ and $j$ with four-momenta $p, p'$ in the center-of-mass frame:
\begin{eqnarray}
{\bf T}_{ij}(p, p') =
 {\bf K}_{ij}(p, p')
 + \sum_n\int{d^4q\over (2\pi)^4} {\bf K}_{in}(p, q)
 \,{\bf G}_n (q)\,{\bf T}_{nj}(q, p')\,\, ~,
\label{T}
\end{eqnarray}
where ${\bf G}$ is the Green function describing the intermediate meson-baryon loop which is iterated to all orders in the integral equation\footnote{Dimensional regularisation with subtraction constants is used in practice.}. The matrix ${\bf K}$ of the driving terms in each channel are constructed from the chiral $SU(3)$ meson-baryon effective Lagrangian in next-to-leading order. In the kaon-nucleon channels, for example, the leading terms have the form\footnote{The convention for the T matrix used here differs from the (dimensionless) one in Ref.\cite{BNW05} by a factor $(2M_N)^{-1}$.}
\begin{equation}
{\bf K}_{K^\pm p} = 2\,{\bf K}_{K^\pm n}  = \mp {\omega\over f^2} + ...\,\, ,
\label{K}
\end{equation}
at zero three-momentum, where the invariant c.m. energy is $\sqrt{s} = \omega + M_N$ (with $M_N$ the nucleon mass) and $f$ is the pseudoscalar meson decay constant. Scattering amplitudes are related to the $T$ matrix (\ref{T}) by ${\bf F} = (M_N/4\pi\sqrt{s})\,{\bf T}$.
Note that ${\bf K} > 0$ means attraction, as seen for example in the $K^- p \rightarrow K^- p$ channel. Similarly, the diagonal matrix elements in the $\pi\Sigma$ channels provide attraction. Close to the $\bar{K}N$ threshold, we have a leading-order piece $F(K^- p \rightarrow K^- p) \simeq (1+m_K/M_N)^{-1}\,m_K/4\pi f^2$. This is the analogue of the Tomozawa-Weinberg term (proportional to $m_\pi/4\pi f^2$ in pion-nucleon scattering at threshold), but now with an attractive strength considerably enhanced by the much larger kaon mass $m_K$.
 
When combining chiral effective field theory with the coupled-channels scheme, the ``rigorous" chiral counting in powers of small momenta is abandoned in favor of iterating a subclass of loop diagrams to ${\it all}$ orders. However, the substantial gain in physics compensates for the sacrifice in the chiral book-keeping.  Important non-perturbative effects are now included in the re-summation, and necessary conditions of unitarity are fulfilled. 

\subsection{S-wave interactions} 
\begin{figure}
\centering
\includegraphics[width=7cm]{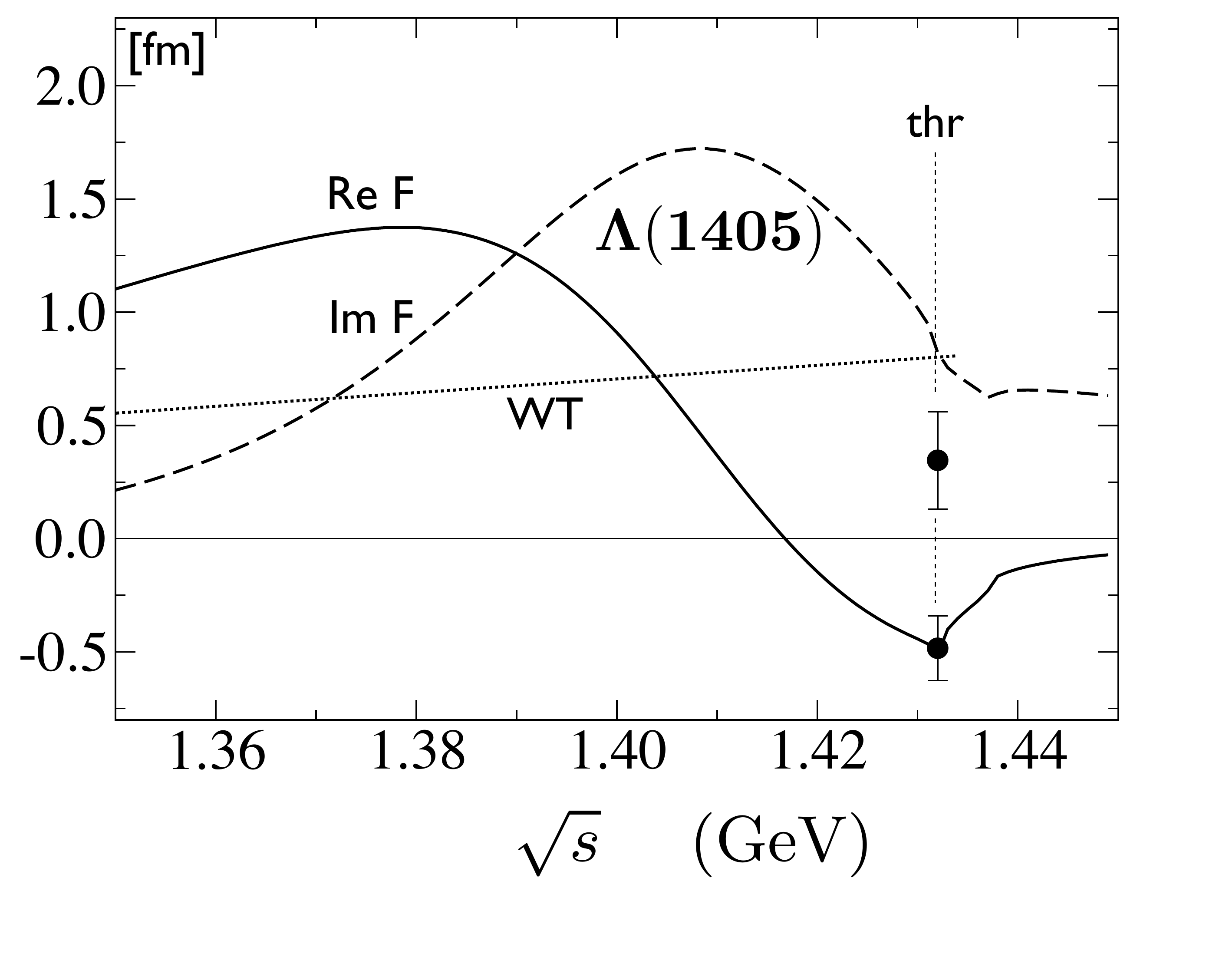}
\caption{Real and imaginary parts of the $K^- p$ forward scattering amplitude calculated in the chiral SU(3) coupled channels approach \cite{BNW05}, as functions of  the invariant $\bar{K}N$ center-of-mass energy$\sqrt{s}$. Real and imaginary parts of the scattering length deduced from the DEAR kaonic hydrogen measurements \cite{Beer04} are also shown. The dotted line indicates the leading-order (Tomozawa-Weinberg) $K^-p\rightarrow K^-p$ amplitude for comparison.}
\label{fig:1}      
\end{figure}

The $K^- p$ threshold data base has recently been improved by new accurate results for the strong interaction shift and width of kaonic hydrogen \cite{Beer04}. These data, together with existing information on $K^- p$ scattering, the $\pi\Sigma$ mass 
spectrum and measured $K^- p$ threshold decay ratios, set tight constraints on the theory and have consequently revived the interest in this field. Fig.\ref{fig:1} shows results of a calculation which combines driving terms from the next-to-leading order chiral $SU(3)$ meson-baryon Lagrangian with coupled-channel equations \cite{BNW05}. As in previous calculations of such kind, the $\Lambda(1405)$ is generated dynamically as an $I = 0$ $\bar{K}N$ quasibound state embedded in the resonant $\pi\Sigma$ continuum. 

The improved accuracy of the kaonic hydrogen data from the DEAR experiment indicate a possible inconsistency with older $K^-p$ scattering data (see Ref. \cite{BNW05}). Note that the real part of the $K^-p$ amplitude, when extrapolated into the subthreshold region below the $\Lambda(1405)$, is expected to be large and positive (attractive).  The imaginary part of this amplitude drops at energies below the $\Lambda(1405)$. The dominant $I = 0$ decay into $\pi\Sigma$ is turned off below its threshold at $\sqrt{s} \simeq 1.33$ GeV. The s-wave $K^-n$ subthreshold amplitude, not shown here, is also attractive but less than half as strong as the $K^-p$ amplitude and non-resonant \cite{BNW05}.  

The off-shell $s$-wave $K^- p$ amplitude resulting from the coupled-channel calculation \cite{BNW05} can be given a convenient approximate parametrisation as follows:
\begin{eqnarray}
F_{K^- p}^{s-wave} = {M_N\over 4\pi f_K^2\sqrt{s}} \left(\omega + a_p\,m_K^2 + b_p\,\omega^2\right) \left(1+{\sqrt{s}\,\gamma_0\over M_0^2 - s -i\sqrt{s}\,\Gamma_0(s)}\right)~~,
\label{swave} 
\end{eqnarray}
with the kaon decay constant $f_K \simeq 0.11$ GeV, $a_b \simeq -b_p \simeq 1$ GeV$^{-1}$, $\gamma_0 \simeq 0.25$ GeV and the $\Lambda(1405)$ mass and energy-dependent width ($M_0, \Gamma_0(s)$), notably with $M_0$ shifted upward by about 10 MeV from its nominal value. This form is useful for practical purposes and reflects the behavior of the leading and next-to-leading order terms as well as the non-perturbative part involving the dynamically produced resonance.  

\subsection{An equivalent pseudopotential}

In applications to nuclear few-body systems it is convenient to translate the leading $\bar{K}N$ $s$-wave interaction into an equivalent potential in the laboratory frame (where the nucleon is approximately at rest). The leading order piece (the Tomozawa-Weinberg term) can be viewed as resulting from vector meson exchange \cite{KW}. Starting from the non-linear sigma model in $SU(3)$, introduce gauge couplings of the vector meson octet to the pseudoscalar octet and fix a universal vector coupling constant $g\simeq 6$ such the $\rho \rightarrow \pi^+\pi^-$ width is reproduced. Then construct vector meson couplings to the $SU(3)$ octet baryons through their conserved vector currents.  The corresponding piece of the reduced $\bar{K}N$ interaction Lagrangian which generates the t-channel vector meson exchange $\bar{K}N$ amplitude at tree level, with vector meson mass $m_V$, is
\begin{eqnarray}
\delta{\cal L}(\bar{K}N)= {ig^2\over 4}\left(K^-\partial_\mu K^+ - K^+\partial_\mu K^-\right)\left[\partial^2 + m_V^2\right]^{-1} \,\bar{\Psi}_N\,\gamma^\mu(\tau_3 + 3)\,\Psi_N~~,
\end{eqnarray}
where $K^{\pm}$ are the charged kaon fields and $\Psi_N = (p,n)^T$ is the isodoublet nucleon field. The isovector ($\tau_3$) piece comes from $\rho$ exchange and the isoscalar part (with its typical factor of 3) comes from $\omega$ exchange, while $\phi$ exchange does not contribute as long as there are no strange quark components in the nucleon. Taking the long-wavelength limit $|\vec{q}\,|\rightarrow 0$, one arrives at the scattering operator 
\begin{eqnarray}
\delta\hat{T} = {g^2\over 2m_V^2}\,\omega\, \Psi_N^\dagger\,(\tau_3 + 3)\,\Psi_N~~.\nonumber
\end{eqnarray}
Using the KSFR relation $m_V = \sqrt{2} f g$, the Tomozawa-Weinberg amplitudes (\ref{K}) follow immediately.

These considerations suggest a characteristic range $r\sim m_V^{-1}$ of the $\bar{K}N$ interaction even for $pointlike$ kaon and nucleon. When the actual size of the nucleon is taken into account,
the minimal range of the s-wave $\bar{K}N$ interaction is determined by the form factors related to the vector currents of the nucleon, for which the electromagnetic form factors of the proton are a good measure. A conservative estimate of this range is therefore given by the r.m.s. proton charge radius range.

The static pseudopotentials which approximate the $\bar{K}N$ interaction in $r$-space follow through the operator identity $\hat{V} = -\delta\hat{T}/2\omega$. The result is:
\begin{equation}
V_{K^-p}(\vec{r}\,) = - {g_p(\vec{r}\,)\over 2f^2}~,~~V_{K^-n}(\vec{r}\,) = - {g_n(\vec{r}\,)\over 4f^2}~~,
\label{pot}
\end{equation}
with distributions $g_{p,n}(\vec{r}\,)$ normalised as $\int d^3r\,g(\vec{r}\,) = 1$. In the limit $m_V\rightarrow\infty$ and for pointlike nucleons,  $g_{p,n}(\vec{r}\,) \rightarrow \delta^3(\vec{r}\,)$. 

Applications in few-body calculations commonly use Gaussian forms for $g(\vec{r}\,)$, with range parameters left free and usually chosen smaller than the r.m.s. radius related to the nucleon's vector current. The reason is that the
effective $\bar{K}N$ interaction, with the $\pi\Sigma$ degrees of freedom ``integrated out", involves iterations of the $\pi\Sigma$ s-channel loops to all orders. Regularisation of these s-channel loops
requires subtraction constants encoding the high-energy behaviour of the loop integrals.  When 
approximating such non-local structures by a local finite-range potential its apparent ``range", treated as a fit parameter, reflects the cutoffs or subtraction constants rather than the range of the driving t-channel
interaction.    

With $f\simeq 0.1$ GeV, the potential $V_{K^-p}(\vec{r}\,)$ is not sufficiently strong to produce a quasibound state. In the work of Ref.\cite{AY02}, the coupling strength was roughly doubled in order to generate the $\Lambda(1405)$, and needed to be even further amplified to deal with the (at that time still recorded) candidates for narrow, deeply bound $K^-NNN$ states. Such a  procedure can be misleading. In fact any approach that tries to generate $K^-$nuclear states using a purely phenomenological static, local and energy-independent $\bar{K}N$ potential misses important physics, for the following reasons.

The $\bar{K}N\leftrightarrow\pi\Sigma$ coupling is well known to be strong. The measured threshold branching ratios for $K^-p$ into $\pi^\pm\Sigma^\mp$ represent about $2/3$ of all $K^-p$ inelastic channels. Moreover, the large fraction of double charge exchange, $\Gamma(K^-p\rightarrow\pi^+\Sigma^-)/\Gamma(K^-p\rightarrow\pi^-\Sigma^+) \simeq 2.4$, demonstrates the importance of coupled-channel dynamics beyond leading orders. An effective potential, projected into the diagonal $K^-p$ channel, that accounts for these mechanisms will necessarily be complex, non-local and energy dependent.  

Consider for simplified demonstration a schematic two-channel model involving the coupled $I=0$ states $|1\rangle = |K^-p\rangle$ and  $|2\rangle = |\pi\Sigma\rangle$. We ignore the (relatively unimportant)
$K^-p\leftrightarrow\bar{K}^0 n$ charge exchange channel and let $ |\pi\Sigma\rangle$ stand for all combinations of charges $\pi^\pm\Sigma^\mp$ and  $\pi^0\Sigma^0$. The channel coupling matrix element $V_{12} = \langle K^-p|V|\pi\Sigma\rangle= V_{21}^*$ is not much weaker than the diagonal elements $V_{11} = \langle K^-p|V|K^-p\rangle$ and $V_{22} = \langle\pi\Sigma|V|\pi\Sigma\rangle$. Let  $h_{1,2}$ include masses and kinetic energies in the respective channels and let the wave function of the coupled system be written $|\psi\rangle = c_1\, |K^-p\rangle +   c_2\, |\pi\Sigma\rangle$:
\begin{eqnarray}
(h_1 + V_{11} - E)\, c_1&=& -V_{12}\, c_2~~,\nonumber \\
(h_2 + V_{22} - E)\, c_2 &=& -V_{21}\, c_1~~.
\label{cc}
\end{eqnarray}
An important part of the mechanism generating the $\Lambda(1405)$ is resonance formation in the $\pi\Sigma$ channel. Assume therefore that the uncoupled equation of motion for $|\pi\Sigma\rangle$ produces a pole at $E = m_0 - (i/2)\Gamma_0$. The $\Lambda(1405)$ with its physical mass $M_0$ is then supposed to emerge as a $K^-p$ (quasi-)bound state embedded in the $\pi\Sigma$ continuum once the channel coupling is turned on. Eliminating $c_2$ from Eq.(\ref{cc}), the remaining equation for $c_1$ projected into the $K^-p$ channel involves the complex and energy dependent effective potential
\begin{equation}
V_{eff}(E) = V_{11} - {|V_{12}|^2\over m_0-{i\over 2}\Gamma_0 -E}~~.
\end{equation} 
Such a non-local $K^-p$ interaction $V_{eff}(E)$ is to be used in variational calculations which do not treat the $\pi\Sigma$ channels explicitly. 

The explicit and detailed construction of an effective single-channel $\bar{K}N$ potential representing $\bar{K}N\leftrightarrow\pi\Sigma$ coupled-channel dynamics, sketched here only in schematic form, has been
carried out systematically in Ref.\cite{HW07}. The result differs significantly from the purely phenomenological potentials used in Refs.\cite{AY02,YA02,YA07}. It turns out that the far subthreshold
extrapolation of the effective interaction based on chiral $SU(3)$ dynamics is in fact substantially less attractive than the phenomenolgical one. 

The reason can be traced to the two-pole structure of the
coupled $\bar{K}N$ and $\pi\Sigma$ amplitudes.  Taken separately, the attractive $ I= 0$ $\bar{K}N$
interaction produces a weakly bound $\bar{K}N$ state, while the attractive $\pi\Sigma$ interaction generates a broad resonance. The coupling of these channels moves the maximum of the $\pi\Sigma$ mass spectrum to 1405 MeV (which is commonly (mis)interpreted as the position of the $\Lambda(1405)$). However, the pole of the $ I =0$ $\bar{K}N$ amplitude turns out to be located at 1420 MeV (see Ref.\cite{HW07} for a detailed discussion). Hence a phenomenological potential describing the physics in the $\bar{K}N$ channel should reproduce the pole at 1420 MeV rather than the ``nominal"  $\Lambda(1405)$. Therefore, given that the ``binding" in the $K^- p$ system is only 12 MeV in this approach \cite{HW07}, less than half of the 27 MeV binding commonly assigned to the $\Lambda(1405)$, it may not be surprising that the binding of the prototype $K^-pp$ cluster turns out weaker \cite{DHW08} than previously suggested. 
 
\subsection{P-wave interactions}

P-waves play a minor role in $\bar{K}N$ interactions close to threshold. However, as pointed out in Ref.\cite{GW05}, they are of potential importance for tightly bound $\bar{K}$-nuclear systems in which the antikaon can have large three-momentum. A useful parametrisation of these amplitudes\footnote{This is an update of the form given long ago in Ref.\cite{BWT78}.} involves dominantly the $\Sigma(1385)$ resonance accompanied by a small background term:
\begin{equation}
F_{K^- p}^{p-wave} = {1\over 2}F_{K^- n}^{p-wave}=  {M_N\over \sqrt{s}}\,C(s)\,\vec{q}\cdot \vec{q}\,' ~~,
\label{p1} 
\end{equation}
where $\vec{q}$ and  $\vec{q}\,'$ are the momenta in the meson-nucleon c.m. frame, and
\begin{equation}
C(s) = {\sqrt{s}\,\gamma_1\over M_1^2 - s -i\sqrt{s}\,\Gamma_1(s)} + d~~,
\label{p2} 
\end{equation}
with $\gamma_1 \simeq 0.42/m_K^2$, $d \simeq 0.06$ fm$^3$, $M_1 = 1.385$ GeV and  the (energy dependent) width  $\Gamma_1(s)$ ($\Gamma_1\simeq 40$ MeV at resonance). Note that these $p$-wave amplitudes
represent attractive interactions below the $\Sigma(1385)$. Here the isospin $I=1$ dominates so that, unlike the $s$-wave case, the $K^-$neutron interaction is now twice as strong as that for $K^-$proton. 

\section{Antikaon-nuclear bound states}

\subsection{The $K^-pp$ system}

Theoretical investigations of prototype $K^-pp$ quasibound states have used two complementary approaches: the variational AMD (Antisymmetrized Molecular Dynamics) method \cite{YA02,YA07} and three-body coupled-channel Faddeev calculations \cite{SGM06,IS07}. 

The Faddeev calculations are performed using separable $NN$, $\bar{K}N$ and $YN$ interactions and include $\bar{K}N \leftrightarrow\pi\Sigma$ channel coupling. The input parameters are constrained by properties of the $\Lambda(1405)$ and by low-energy scattering data where available.
The calculated pole positions  \cite{SGM06,IS07} of the three-body T-matrix in the complex plane give the following range for binding energy and width of $K^-pp$:
\begin{eqnarray}
B(K^-pp) &\sim& (55 - 80)\,{\mbox{MeV}}\,,\nonumber \\
\Gamma(K^-pp\rightarrow\pi\Sigma N) &\sim& (75 - 110)\, {\mbox{MeV}}~, 
\end{eqnarray}
depending on details of the input parameter sets which determine the $\bar{K}N$ interaction. The
largest $B$ and lowest value of $\Gamma$ is actually found with 
the separable approximation to a chiral $SU(3)$ based $\bar{K}N$ interaction used in Ref.\cite{IS07}.
Questions about ambiguities in the far subthreshold, off-shell extrapolation remain for all such interactions. 
    
The absorptive width $\Gamma(K^-pp\rightarrow YN)$, not included in these computations, would add to increase the total width well beyond 100 MeV. While these first exploratory variational and Faddeev calculations are roughly consistent amongst themselves, they are (so far) not compatible with the interpretation of the FINUDA data \cite{Ag05} as signals for the formation of deeply bound $K^-pp$ clusters with binding energy as large as $B(K^-pp) \sim 115$ MeV and a width around 70 MeV.

Recent variational calculations  \cite{DHW08} using the chiral $SU(3)$ effective $\bar{K}N$ interaction derived in Ref.\cite{HW07} and mentioned previously, suggest instead weaker $K^-pp$ binding, with a
binding energy window between 15 and 25 MeV and widths in the range 40 - 60 MeV (not including
$K^-pp\rightarrow YN$ absorption). Such states would be practically undetectable.
 
\subsection{Antikaons in nuclear matter}

Kaonic nuclei with a $K^-$ bound to heavier nuclear cores are likewise of interest even though their experimental detection would certainly be difficult. As a generic starting point of this discussion,
consider $K^-$ modes in nuclear matter. The kaon spectrum in matter with proton and neutron densities $\rho_{p,n}$ is determined by
\begin{equation}
\omega^2 - \vec{q}\,^2 - m_K^2 - \Pi_K(\omega,\vec{q}\,;\rho_{p,n}) = 0~~,
\label{Keqn}
\end{equation} 
with the $K^-$ self-energy $\Pi_K$ (or equivalently, the $K^-$ nuclear potential $U_K$) in the nuclear medium: 
\begin{equation}
\Pi_{K^-} = 2\omega \,U_{K^-} = -T_{K^-p}\,\rho_p - T_{K^-n}\,\rho_n + \, ...
\end{equation} 
to leading order in the nucleon densities, where $T_{K^-p,n}$ are the $K^-p$ and $K^-n$ (forward scattering) T-matrices. The additional terms, not shown explicitly, include corrections from Fermi motion, Pauli blocking, two-nucleon correlations etc.
An effective kaon mass in the medium can be introduced by solving Eq.(\ref{Keqn}) at zero momentum: 
\begin{eqnarray}
m_K^*(\rho) = \omega(\vec{q} = 0, \rho)~~.
\nonumber
\end{eqnarray}

\begin{figure}
\centering
\includegraphics[width=7.5cm]{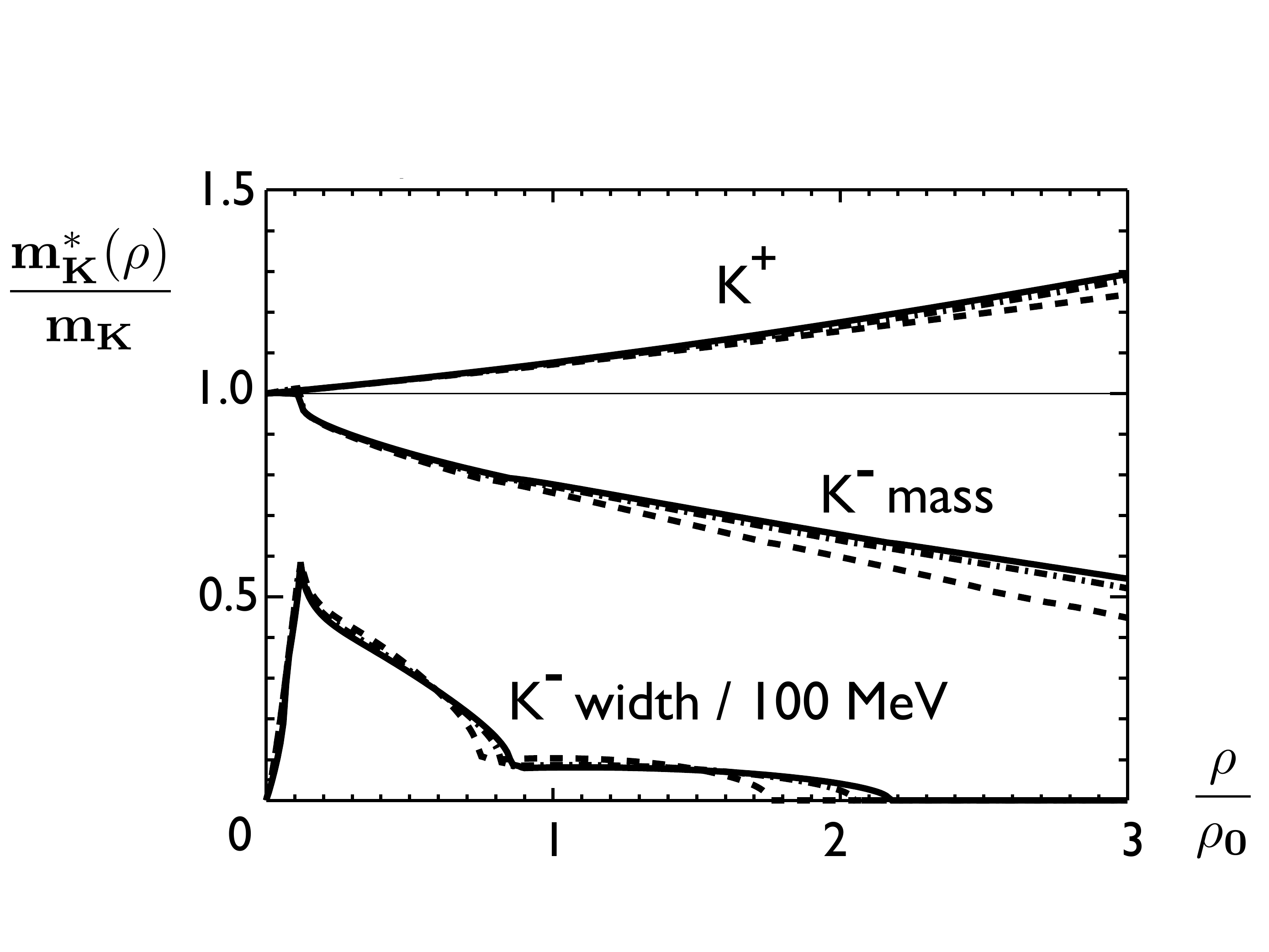}
\caption{In-medium mass $m_K^*(\rho)$ and width of a $K^-$ in symmetric nuclear matter as a function of baryon density $\rho$ in units of nuclear matter density $\rho_0 = 0.17$ fm$^{-3}$. The calculations \cite{WKW96} were performed using in-medium chiral SU(3) dynamics combined with coupled channels and including effects of Pauli blocking, Fermi motion and two-nucleon correlations. Also shown is the in-medium $K^+$ effective mass calculated in the same approach.}
\label{fig:2}      
\end{figure}

Calculations of the spectrum of kaonic modes as a function of density have already a long history. For example, in Refs. \cite{WKW96} it was pointed out that, as a consequence of the underlying attractive $\bar{K}N$ forces, the $K^-$ mass at the density of normal nuclear matter ($\rho_0 \simeq 0.17$ fm$^{-3}$) effectively drops to about three quarters of its vacuum value. At this density the $K^-$ in-medium decay width is expected to be strongly reduced because the $K^- N$ energy ``at rest" in matter has already fallen below the $\pi\Sigma$ threshold. These calculations do, however, not include the $\bar{K}NN\rightarrow YN$ absorptive width $\Gamma_{abs}$ which grows with $\rho^2$, the square of the baryonic density. A rough estimate (see Section \ref{Knuclei}) gives $\Gamma_{abs} \sim 30$ MeV at $\rho = \rho_0$ which adds to the width shown in Fig.\ref{fig:2}. 

\subsection{Kaonic nuclei}
\label{Knuclei}
Mares et al. \cite{MFG06} have studied the possiblity of $\bar{K}$-nuclear bound states using a relativistic mean field model in which the $\bar{K}$ couples to scalar and vector fields mediating the nuclear interactions. Estimates of the absorptive width are also made. Kaon-nuclear binding energies are found in the range $B_K \sim 100 - 200$ MeV accompanied by widths with a lower limit of about 50 MeV.

An alternative, exploratory calculation, using a realistic subthreshold $\bar{K}N$ interaction as described in Section 2, can be based on the Klein-Gordon equation with a complex, energy dependent $K^-$ self-energy $\Pi_K(\omega, \vec{r}\,)$. Bound states are determined as eigenstates of

\begin{equation}
[\omega^2 + \vec{\nabla}^2 - m_K^2 - {\mbox{Re}}\,\Pi_K(\omega,\vec{r}\,)]\,\phi_K(\vec{r}\,) = 0~~.
\label{KGeqn}
\end{equation}
The local Coulomb potential $V_c(r)$ is consistently introduced by the gauge invariant replacement $\omega \rightarrow \omega - V_c(r)$ and understood to be incorporated in all subsequent steps. The width of the bound state is calculated according to
\begin{equation}
\Gamma = -\int d^3r \,\phi_K^*(\vec{r}\,)\, {{\mbox{Im}}\,\Pi_K\over \omega}\,\phi_K(\vec{r}\,)~~.
\label{width}
\end{equation} 
The kaon self-energy includes $s$- and $p$-wave interactions to leading order in density:
\begin{eqnarray}
\Pi_K(\omega, \vec{r}\,) &=& \Pi_s(\omega, \vec{r}\,) + \Pi_p(\omega, \vec{r}\,) + \Delta\Pi_K~~,\\
\Pi_s(\omega, \vec{r}\,) &=& - 4\pi\left(1 + {\omega\over M_N}\right)\left[ F_{K^-p}(\omega)\,\rho_p(\vec r\,)+  F_{K^-n}(\omega)\,\rho_n(\vec r\,)\right] ~~,\nonumber\\
\Pi_p(\omega, \vec{r}\,) &=& 4\pi\left(1 + {\omega\over M_N}\right)^{-1}\vec{\nabla}\left[ C_{K^-p}(\omega)\,\rho_p(\vec r\,)+  C_{K^-n}(\omega)\,\rho_n(\vec r\,)\right]\cdot\vec{\nabla} ~~.\nonumber
\end{eqnarray} 
The $s$-wave amplitudes, $F_{K^-p}(\omega)$ and $F_{K^-n}(\omega)$, are taken from Ref.\cite{BNW05} and handled in parametrised form (see e.g. Eq.(\ref{swave})). The $p$-wave amplitudes, $C(\omega)$, are given by Eqs.(\ref{p1}, \ref{p2}). The term $\Delta\Pi_K$ stands for a series of higher-order corrections (Pauli and short-range correlations, two-nucleon absorption etc.). Pauli and short-range correlations including charge exchange channels and a Lorentz-Lorenz correction for the $p$-wave parts  are dealt with using the methods explained in detail in Ref.\cite{WRW97} for nuclear matter, but now transcribed using local density distributions. 

The proton and neutron densities $\rho_p(\vec{r}\,) = \rho_0 (Z/A) w_p(\vec{r}\,)$ and $\rho_n(\vec{r}\,) = \rho_0 (N/A) w_n(\vec{r}\,)$ are parametrized in terms of Woods-Saxon type distributions $w(\vec{r}\,)$ normalized to unity. The central density $\rho_0$ is varied in order to examine the effects of a possible compression of the core nuclei. 

The influence of two-nucleon absorption processes on the bound state widths is estimated introducing an absorptive piece 

\begin{equation}
\Delta\Pi_{abs} = -4\pi i\,B_0\left(1+{\omega\over 2M_N}\right)\eta(\omega)\,\rho^2(r)~~,
\label{abs}
\end{equation}
with
\begin{equation}
\eta(\omega) = {1\over A^2}\left(Z(Z-1)\,\beta_{pp}(\omega) + 2ZN\,\beta_{pn}(\omega)+ {1\over 3}N(N-1)\,\beta_{nn}(\omega)\right)~~.
\label{abs}
\end{equation}
The reduced $\rho_n^2$ term approximately takes into account the fact that $K^-$ absorption on a neutron pair can only lead to a single $\Sigma^-n$ final state whereas absorption on $pp$ and $pn$ pairs generates $\Sigma N$ and $\Lambda N$ with a greater variety of charge combinations. The kinematical factors $\beta_{ij}(\omega)$, normalised to unity at $\omega = m_K$, take in account the phase space dependence in the respective $K^-NN\rightarrow YN$  channels. The constant $B_0$ is subject to large uncertainties. For orientation we use $B_0 \sim 1$ fm$^4$ guided by constraints from the widths of kaonic atom states \cite{FG07}, but we also allow for a range of values between 0.85 and 1.5 $fm^4$ in order to give conservative error estimates.

\begin{figure}[htb]
\begin{minipage}[t]{65mm}
\includegraphics[width=6.5cm]{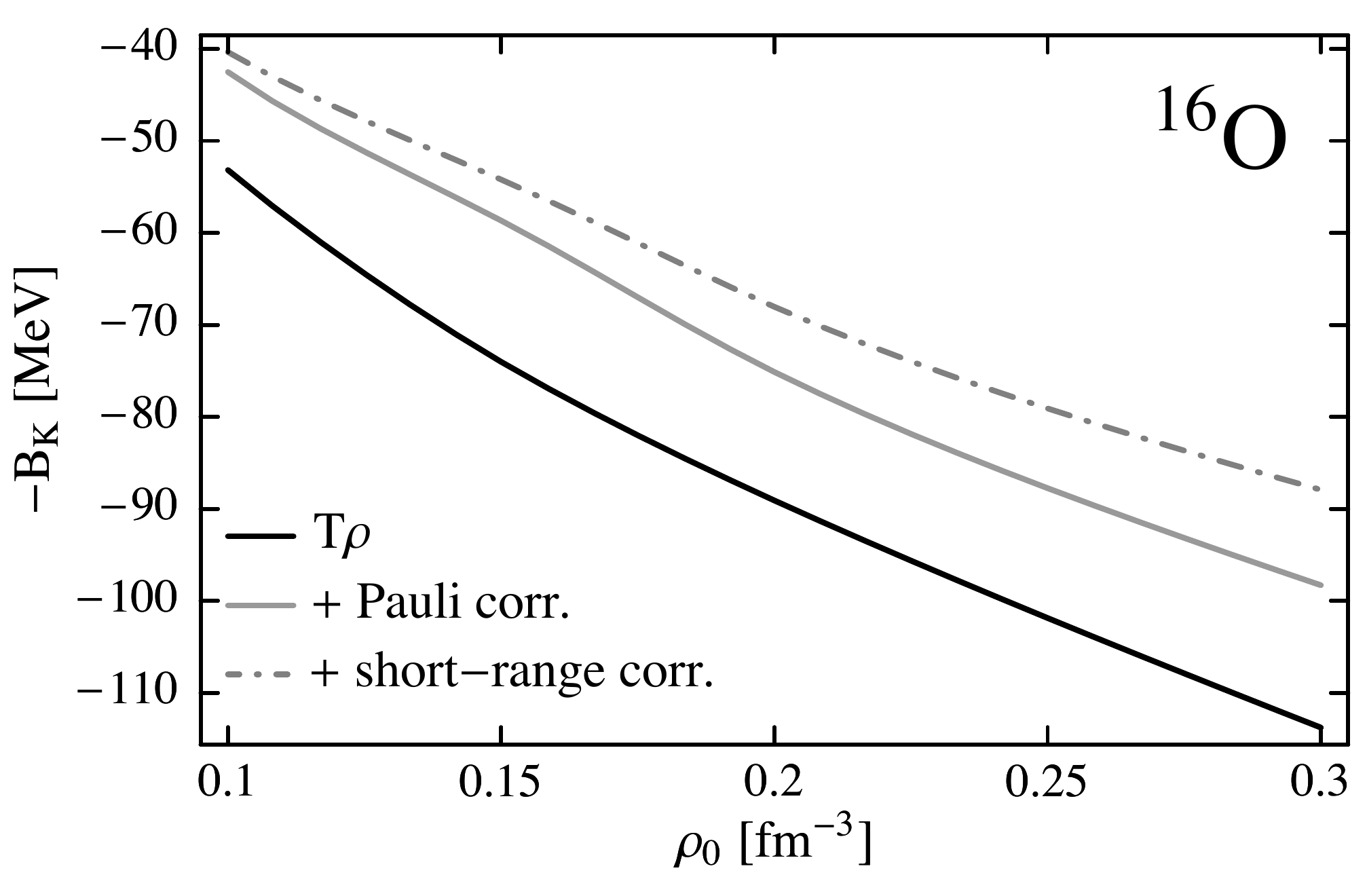}
\caption{Binding energy (plotted as $-B_K$) of $K^-$ bound in $^{16}O$ as function of the central nuclear density $\rho_0$. The solid dark curve results from leading order $s$- and $p$-wave interactions ($T\rho$). Curves in light print show the effects of Pauli and short-range NN correlations as indicated.}

\label{fig:3a}
\end{minipage}
\hspace{\fill}
\begin{minipage}[t]{65mm}
\includegraphics[width=6.5cm]{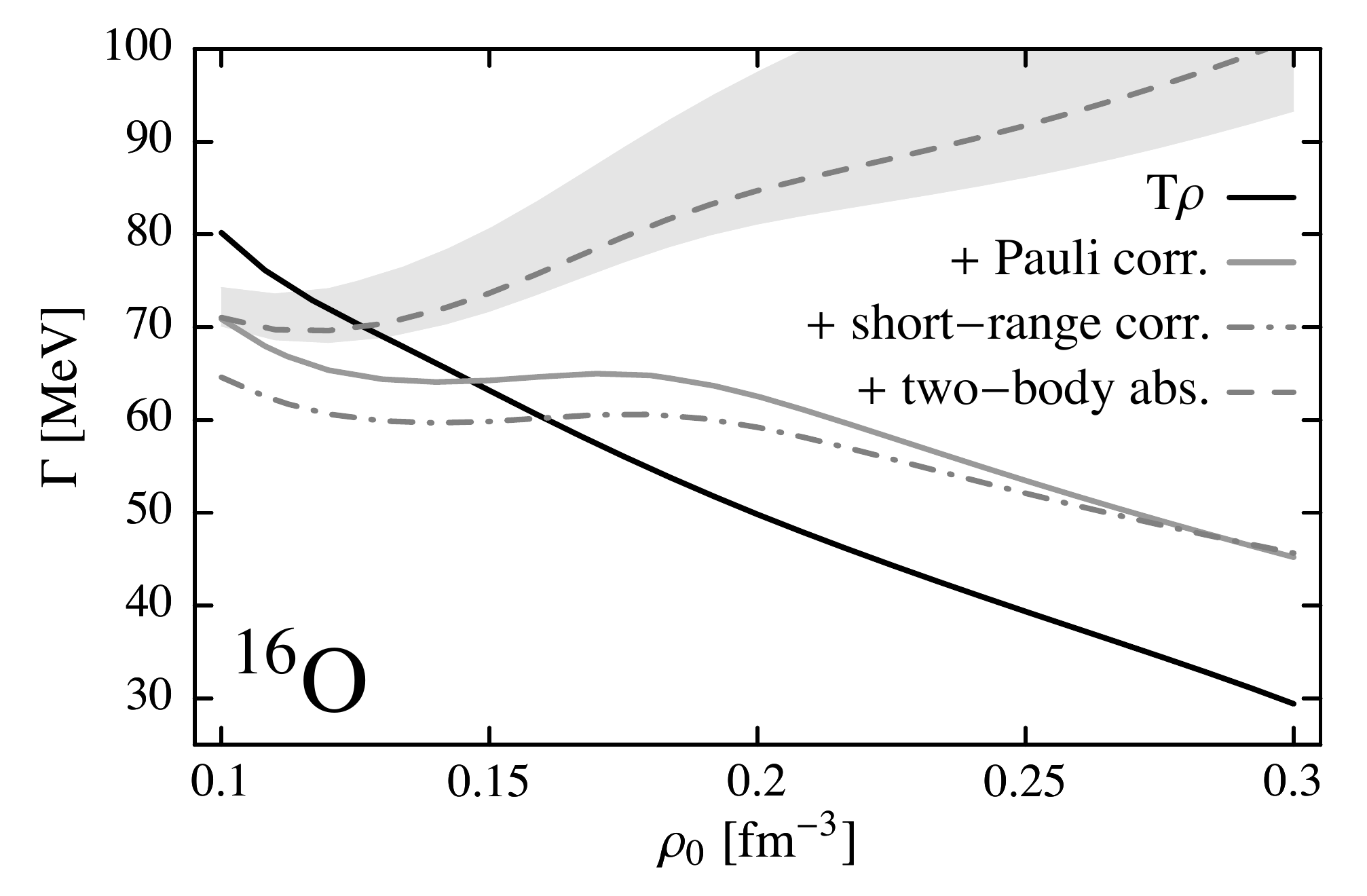}
\caption{Width $\Gamma$ of $K^-$ quasibound state in $^{16}O$ as function of the central nuclear density $\rho_0$, calculated according to Eq.(\ref{width}). The solid dark curve shows the width in $T\rho$ approximation. Other curves show the effects of Pauli and short-range NN correlations as indicated. The dashed curve displays the increased width when $K^-NN\rightarrow YN$ absorption is incorporated, with $B_0 =$ 1 fm$^4$ in Eq.(\ref{abs}); the shaded area gives an impression of the uncertainty range, with $B_0 \in[0.85, 1.5]$ fm$^4$.}
\label{fig:3b}
\end{minipage}
\end{figure}

\begin{figure}[htb]
\begin{minipage}[t]{65mm}
\includegraphics[width=6.5cm]{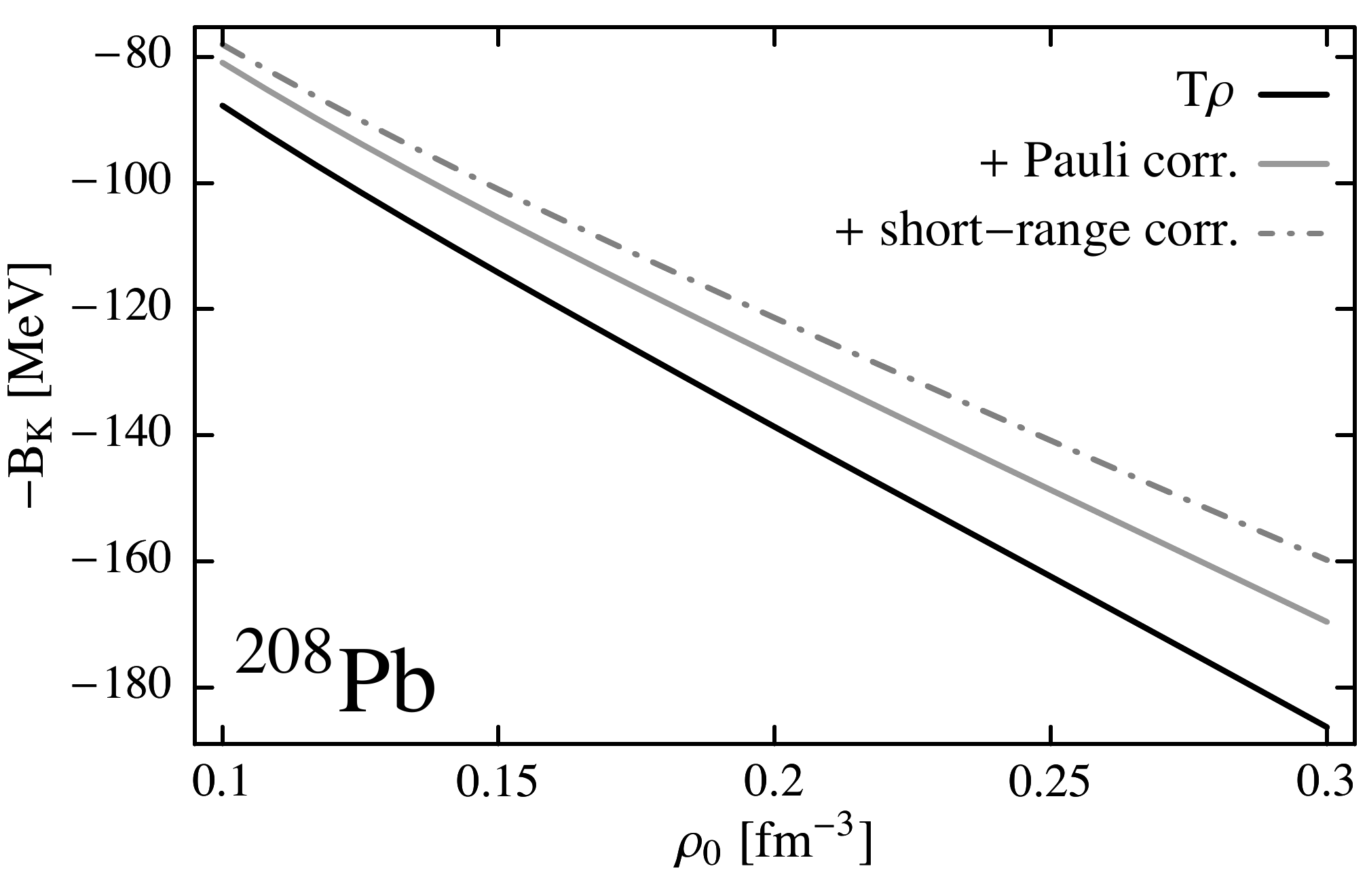}
\caption{Same as Fig. \ref{fig:3a}, for the case of $^{208}Pb$.}
\label{fig:4a}
\end{minipage}
\hspace{\fill}
\begin{minipage}[t]{65mm}
\includegraphics[width=6.5cm]{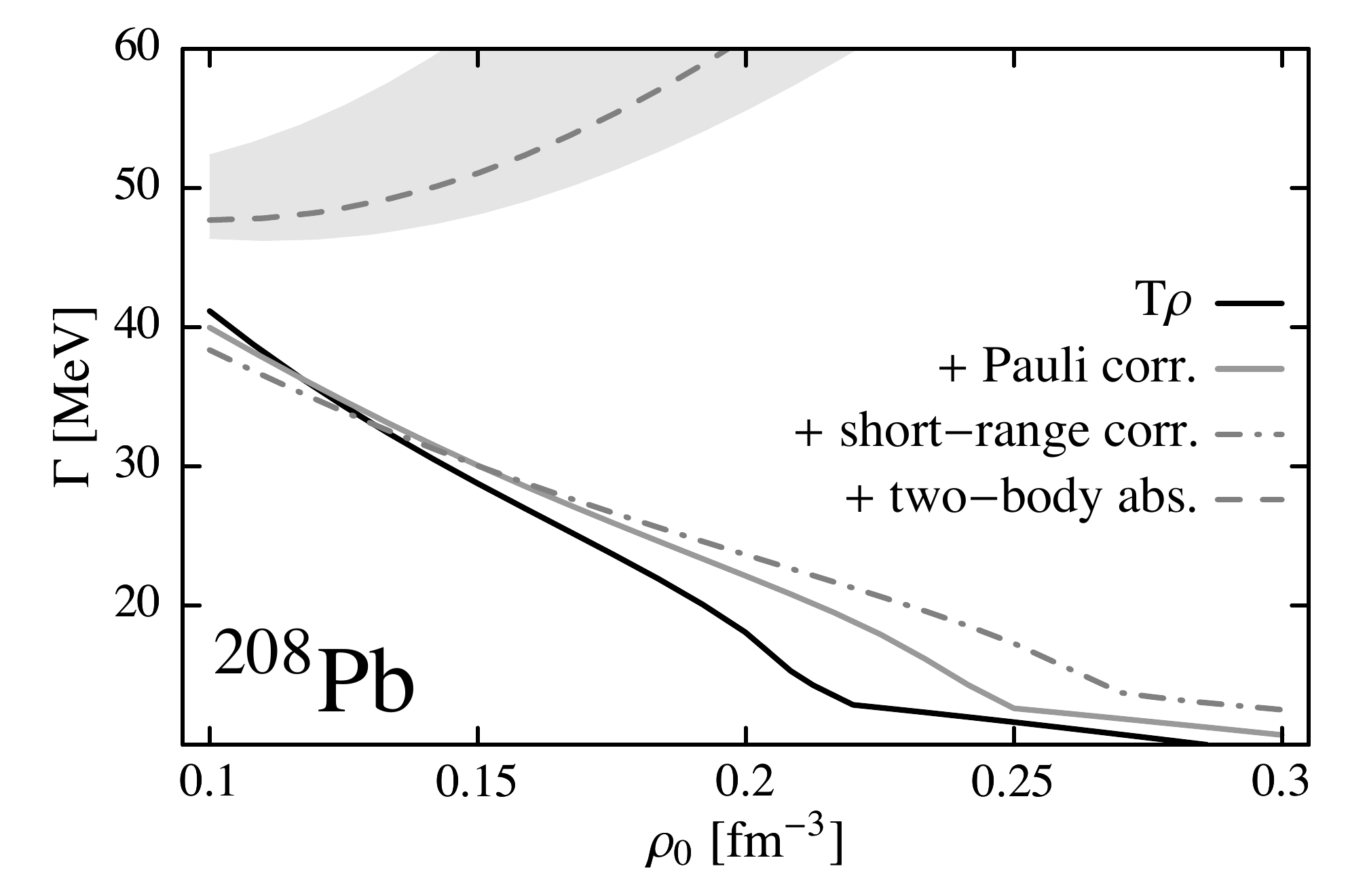}
\caption{Same as Fig. \ref{fig:3b}, for the case of $^{208}Pb$.}
\label{fig:4b}
\end{minipage}
\end{figure}

Representative results \cite{H06} for kaonic nuclei with a $K^-$ bound in $^{16}O$ and $^{208}Pb$ are shown in Figs. \ref{fig:3a}-\ref{fig:4b}. The $K^-$ binding energies for these finite nuclei are calculated as
\begin{equation}
B_K = -{\omega^2 - m_K^2\over 2\omega}\left(1 + {\omega\over M_A}\right)~~,
\label{binden}
\end{equation}
$M_A$ being the mass of the nucleus.

 The leading $s$-wave interaction produces strong antikaon binding. The $p$-wave interaction tends to increase the binding only marginally for $^{208}Pb$ but has a more pronounced effect in lighter nuclei. Pauli and short-range repulsive correlations reduce the binding, as expected. The Coulomb interaction is included in all cases. For $^{208}Pb$, the attractive Coulomb potential experienced by the $K^-$ contributes about 25 MeV to its binding energy at standard nuclear central density, $\rho_0 = 0.17$ fm$^{-3}$.  
 
 The widths are strongly enhanced with increasing density when $K^-NN \rightarrow YN$ absorption is incorporated. This enhancement of the width is more pronounced than in \cite{MFG06} where the absorption term was parametrized as linear (rather than quadratic) in the density; see however Ref.\cite{GF07}.
 
 Note that the effects of Pauli and short-range correlations on the $K^-N\rightarrow\pi\Sigma$ decay widths of quasibound antikaon-nuclear states are relatively small around standard central densities,
 $\rho_0\simeq 0.17$ fm$^{-3}$. At higher densities the reduced binding generated by the repulsive correlations implies an increased phase space for the decay into $\pi\Sigma$ and therefore a moderately incresed width. In any case, at least for heavier nuclei, a large part of the total width of $\bar{K}$-nuclear quasibaound states is expected from  the $K^-NN\rightarrow YN$ absorption processes. 

\begin{figure}[htb]
\begin{minipage}[t]{65mm}
\includegraphics[width=6.5cm]{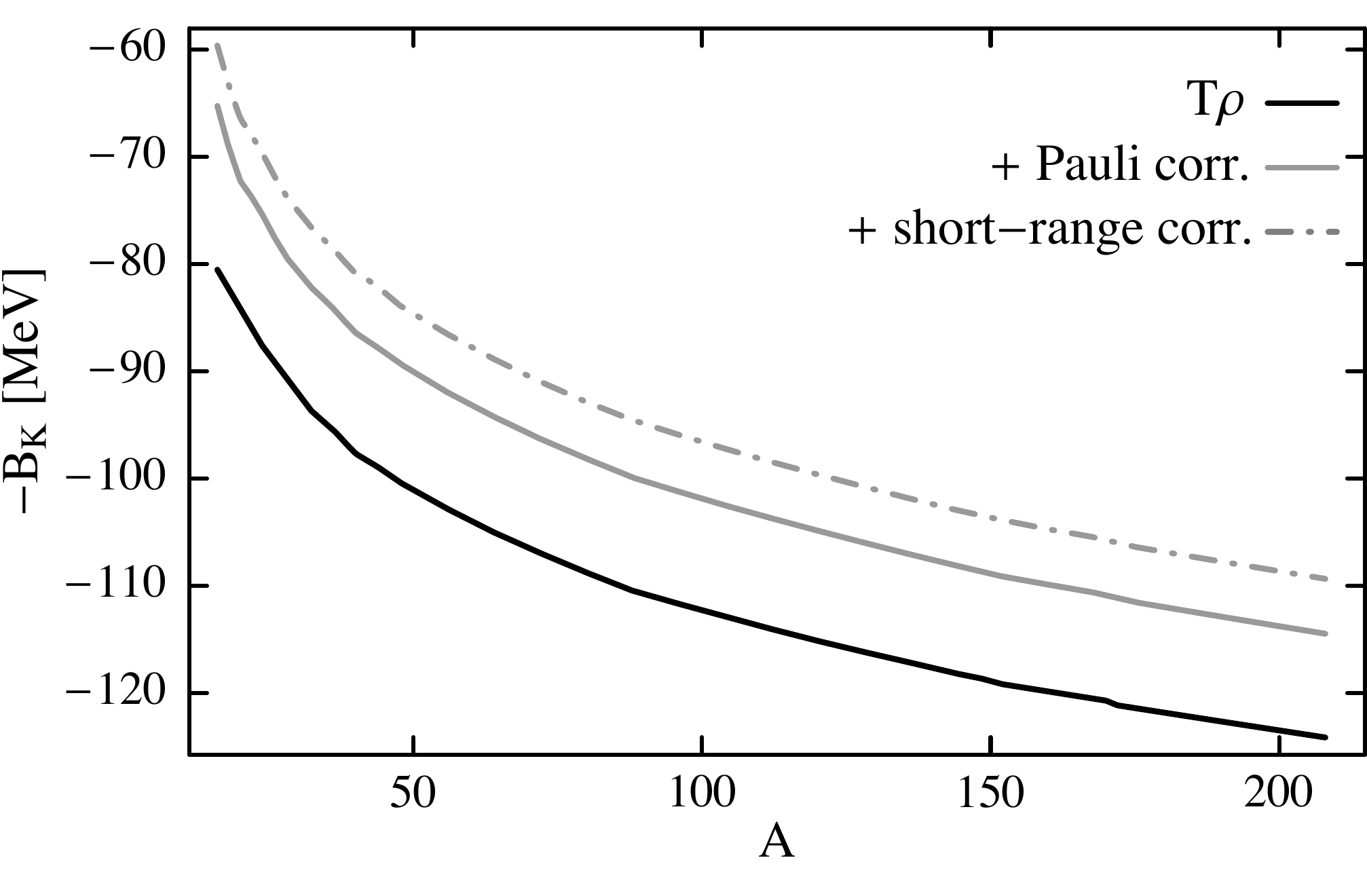}
\caption{Binding energy $B_K$ as a function of nuclear  mass number $A$ at fixed central density
$\rho_0 = 0.17$ fm$^{-3}$. Legends of curves as in Fig. \ref{fig:3a}.}

\label{fig:5a}
\end{minipage}
\hspace{\fill}
\begin{minipage}[t]{65mm}
\includegraphics[width=6.4cm]{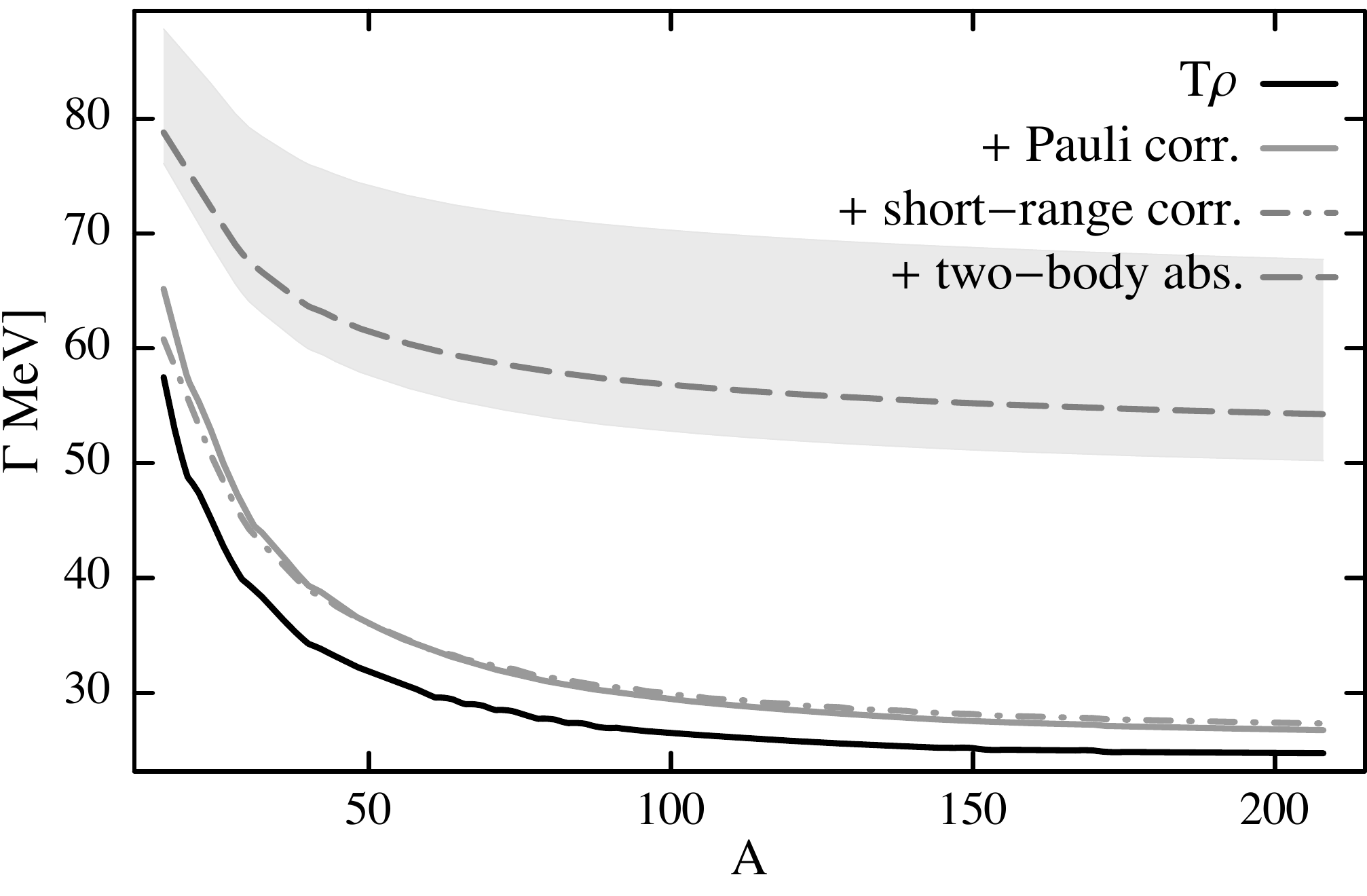}
\caption{Width $\Gamma$ as a function of nuclear  mass number $A$ at fixed central density
$\rho_0 = 0.17$ fm$^{-3}$. Legends of curves as in Fig. \ref{fig:3b}.}
\label{fig:5b}
\end{minipage}
\end{figure}

Figs. \ref{fig:5a}-\ref{fig:5b} show the antikaon-nuclear binding energies and widths  as a function of nuclear mass number, with the central density $\rho_0$ held fixed. The overall systematics seen here is qualitatively similar to the findings in Ref.\cite{MFG06}, although differences in detail result from the different interactions used as input. 

\begin{figure}[htb]
\begin{minipage}[t]{65mm}
\includegraphics[width=6.5cm]{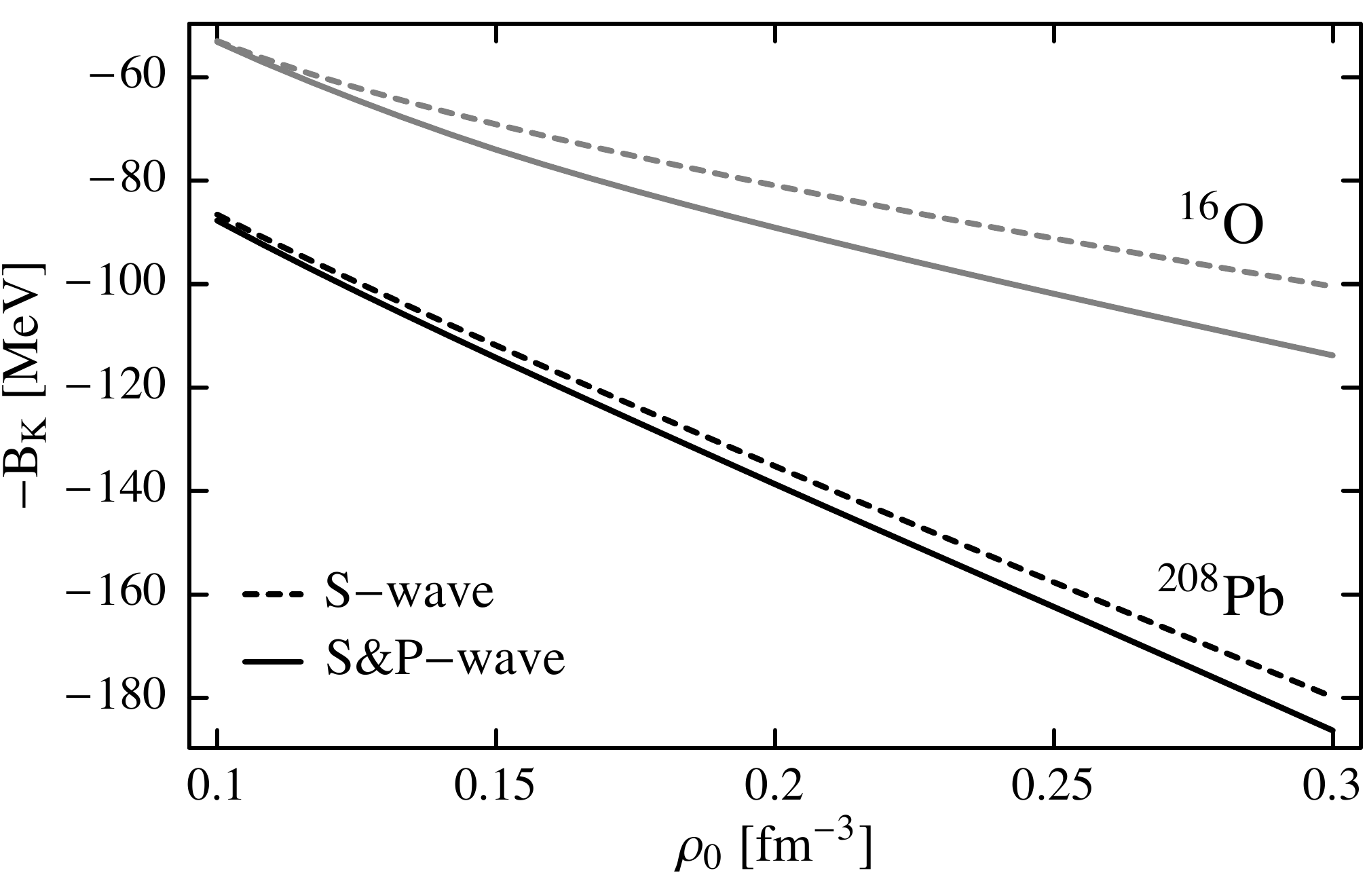}
\caption{Effect of $p$-wave interactions on the binding energies $B_K$ of $^{16}_K O$ and  $^{208}_K Pb$ as function of the central density $\rho_0$.}

\label{fig:6a}
\end{minipage}
\hspace{\fill}
\begin{minipage}[t]{65mm}
\includegraphics[width=6.4cm]{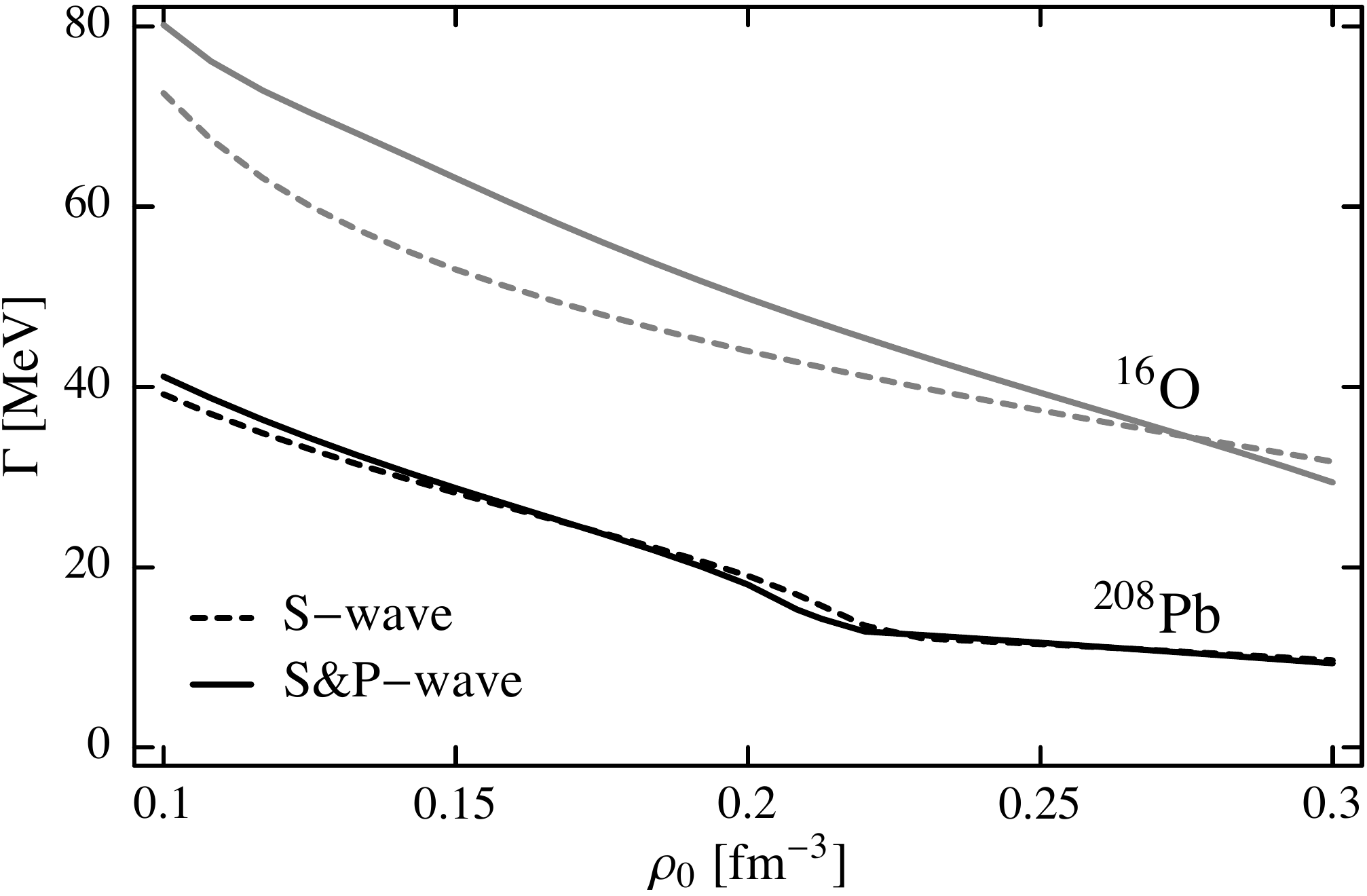}
\caption{Effect of $p$-wave interactions on the widths of $^{16}_K O$ and  $^{208}_K Pb$ as function of the central density $\rho_0$.}
\label{fig:6b}
\end{minipage}
\end{figure}

Finally, the relatively small influence of the $p$-wave $\bar{K}N$ interactions involving the $\Sigma(1385)$ is demonstrated in Figs. \ref{fig:6a}-\ref{fig:6b}. The effects are more pronounced in light nuclei with their larger ratio of surface to bulk, but never dominant. These results are in agreement with Ref.\cite{GF07}, but at variance with discussions in Ref.\cite{GW05}.

\section{Concluding remarks}

The issue of deeply bound antikaon-nuclear systems (``kaonic nuclei") is a very interesting one but so far unsettled. Early model calculations of kaonic few-nucleon systems did not yet use realistic $\bar{K}N$ and $NN$ interactions. More recent computations with improved interactions come to the (tentative) conclusion that $K^-pp$ as a prototype of an antikaon-nuclear cluster is not as deeply bound as anticipated and presumably has a very short lifetime. Previously published narrow $K^-NNN$ signals at KEK have disappeared in a measurement with much improved statistics. The question about the existence of $K^-pp$ and $K^-ppn$ quasibound clusters has been raised independently by the FINUDA measurements but their interpretation is under dispute. Deeply bound $K^-$ states in heavier nuclei may exist, but with large widths. A more detailed understanding of these widths  and their underlying mechanisms calls for systematic, exclusive and kinematically complete measurements of the final states resulting from $K^-$ induced processes, especially in light nuclei.  

\vspace{0.8cm}
\noindent
{\bf Acknowledgements}\\
\\
\noindent
Stimulating discussions with Akinobu Dot$\acute{\rm e}$, Avraham Gal, Tetsuo Hyodo and Paul Kienle are gratefully acknowledged.
This work is supported in part by BMBF, GSI, and by the DFG cluster of excellence Origin and Structure of the Universe.


\begin{thebibliography}{00}
%

\bibitem{KapNel}
D. Kaplan and A. Nelson, Phys. Lett. \textbf{B175} (1986) 57; Nucl. Phys. \textbf{A479} (1988) 273. 

\bibitem{Wy86}
S. Wycech, Nucl. Phys. \textbf{A450} (1986) 399c. 

\bibitem{BLRT94}
G.E. Brown, C.H. Lee, M. Rho and V. Thorsson, Nucl. Phys. \textbf{A567} (1994) 937. 

\bibitem{WKW96} 
T. Waas, N. Kaiser and W. Weise, Phys. Lett. \textbf{B379} (1996) 34;  Phys. Lett. \textbf{B365} (1996) 12; \\T. Waas and W. Weise, Nucl Phys. \textbf{A625} (1997) 287.

\bibitem{WRW97} 
T. Waas, M. Rho and W. Weise, Nucl Phys. \textbf{A617} (1997) 449.

\bibitem{AY02} 
Y. Akaishi and T. Yamazaki, Phys. Rev. \textbf{C65} (2002) 044005; A. Dot$\acute{\rm e}$, H. Horiuchi, Y. Akaishi and T. Yamazaki, Phys. Rev. {\bf C70} (2004) 044313; A. Dot$\acute{\rm e}$, Y. Akaishi and T. Yamazaki, Prog. Theor. Phys. Suppl. \textbf{156} (2004) 184.

\bibitem{YA02}
T. Yamazaki and Y. Akaishi, Phys. Lett. {\bf B535} (2002) 70.

\bibitem{Su04} 
T. Suzuki et al., Phys. Lett. \textbf{B597} (2004) 263.

\bibitem{Iwa06} 
M. Sato {\em et~al.}, arXiv:0708.2968 [nucl-ex]; Phys Lett. {\bf B} (2008), in print.

\bibitem{Ag05} 
M. Agnello et al., Phys. Rev. Lett. \textbf{94} (2005) 212303.

\bibitem{OT06} 
V.K. Magas, A. Ramos, E. Oset and H. Toki,  Phys. Rev. \textbf{C74} (2006) 025206.

\bibitem{Ag07} 
M. Agnello et al., Phys. Lett. \textbf{B654} (2007) 80.

\bibitem{GSI06} 
N. Herrmann et al., in: Proc. Int. Conf. ``Exotic Atoms and Related Topics (EXA05)'', Austrian Acad. of Sc. Press (2006), pp. 73-81. 

\bibitem{SGM06} 
N.V. Shevchenko, A. Gal and J. Mares, Phys. Rev. Lett. \textbf{98} (2007) 082301;\\
N.V. Shevchenko, A. Gal, J. Mares, and J. Revai,  Phys. Rev.  \textbf{C76} (2007) 044004.

\bibitem{IS07}
Y. Ikeda and T. Sato, Phys. Rev. {\bf C76} (2007) 035203. 

\bibitem{DW07} 
A. Dot$\acute{\rm e}$ and W. Weise, Prog. Theor. Phys. Suppl. \textbf{168} (2007) 593 [nucl-th/0701050].  

\bibitem{YA07}
T. Yamazaki and Y. Akaishi, Phys. Rev. {\bf C76} (2007) 04520.

\bibitem{HW07} 
T. Hyodo and W. Weise, arXiv:0712.1613 [nucl-th] (2007).  

\bibitem{DHW08} 
A. Dot$\acute{\rm e}$, T. Hyodo and W. Weise, preprint (2008).  

\bibitem{MFG06} 
J. Mares, E. Friedman and A. Gal, Nucl. Phys. \textbf{A770} (2006) 84.

\bibitem{GF07} 
D. Gazda, E. Friedman, A. Gal and J. Mares, Phys. Rev. \textbf{C76} (2007) 055204.

\bibitem{FG07} 
E. Friedman and A. Gal, Phys. Reports \textbf{452} (2007) 89.

\bibitem{BNW05} 
B. Borasoy, R. Ni{\ss}ler and W. Weise, Phys. Rev. Lett. \textbf{94} (2005) 213401; \\Eur. Phys. J. \textbf{A25} (2005) 79; Phys. Rev. Lett. \textbf{96} (2006) 199201. 

\bibitem{KSW95}
N. Kaiser, P.B. Siegel and W. Weise, Nucl. Phys. \textbf{A594} (1995) 325.

\bibitem{KWW97} 
N. Kaiser, T. Waas and W. Weise, Nucl. Phys.  \textbf{A612} (1997) 297; \\E. Oset and A. Ramos, Nucl. Phys. \textbf{A635} (1998) 99; \\J. Caro Ramon, N. Kaiser, S. Wetzel and W. Weise, Nucl. Phys.  
\textbf{A672} (2000) 249; \\J.A. Oller and U.-G. Mei{\ss}ner, Phys. Lett.  \textbf{B500} (2001) 263; \\M.F.M. Lutz and E. Kolomeitsev, Nucl. Phys.  \textbf{A700} (2002) 193.

\bibitem{BMN06} 
B. Borasoy, U.-G. Mei{\ss}ner and R. Ni{\ss}ler, Phys. Rev. \textbf{C74} (2006) 055201.

\bibitem{Beer04} 
DEAR Collaboration, G. Beer et al., Phys. Rev. Lett. \textbf{94} (2005) 212303.

\bibitem{KW}
F. Klingl, N. Kaiser and Weise, Z. Phys. \textbf{A356} (1996) 193.

\bibitem{GW05} 
S. Wycech and A.M. Green, [nucl-th/0501019].

\bibitem{BWT78} 
R. Brockmann, W. Weise and L. Tauscher, Nucl Phys. \textbf{A308} (1978) 365.

\bibitem{H06} 
R. H\"artle, Diploma Thesis, TU Munich (2006).

\end{thebibliography}
\end{document}